\documentclass[journal]{IEEEtran}

\usepackage[ruled]{algorithm2e}
\usepackage{graphicx, amsfonts}
\usepackage{amsmath, amssymb}
\usepackage{cite, array, multirow}
\usepackage{siunitx}
\usepackage{xcolor, color, hhline, stfloats}
\usepackage{lipsum}
\usepackage{gensymb}
\usepackage{verbatim, booktabs}
\usepackage{algpseudocode}
\usepackage{amsmath}
\usepackage{graphics}
\usepackage{epsfig}
\ifCLASSINFOpdf

\begin{document}
%
\title{Multi-modal Multi-channel Target Speech Separation}
%
%
%

\author{Rongzhi~Gu, ~\IEEEmembership{Student Member,~IEEE},
        Shi-Xiong~Zhang,
        Yong~Xu,
        Lianwu~Chen,\\
        Yuexian~Zou,~\IEEEmembership{Senior Member,~IEEE},
        and~Dong~Yu,~\IEEEmembership{Fellow,~IEEE}
\thanks{Rongzhi Gu and Yuexian Zou are with the Advanced data and signal processing laboratory, School of Electric and Computer Science, Peking University Shenzhen Graduate School, Shenzhen, China. Yuexian Zou is also with Pengcheng Laboratory, Shenzhen, China. This work was done when Rongzhi Gu was an intern at Tencent AI Lab, Shenzhen, China. e-mail: zouyx@pku.edu.cn.}
\thanks{Shi-Xiong Zhang, Yong Xu and Dong Yu are with Tencent AI Lab, Seattle, WA, USA.}
\thanks{Lianwu Chen is with Tencent AI Lab, Shenzhen, China.}}

%
%

\markboth{IEEE Journal of Selected Topics in Signal Processing, 2020}%
{Shell \MakeLowercase{\textit{et al.}}: Bare Demo of IEEEtran.cls for IEEE Journals}
%


\maketitle

\begin{abstract}
Target speech separation refers to extracting a target speaker's voice from an overlapped audio of simultaneous talkers.
Previously the use of visual modality for target speech separation has demonstrated great potentials. This work proposes a general multi-modal framework for target speech separation by utilizing all the available information of the target speaker, including his/her spatial location, voice characteristics and lip movements. Also, under this framework, we investigate on the fusion methods for multi-modal joint modeling. A factorized attention-based fusion method is proposed to aggregate the high-level semantic information of multi-modalities at embedding level. This method firstly factorizes the mixture audio into a set of acoustic subspaces, then leverages the target's information from other modalities to enhance these subspace acoustic embeddings with a learnable attention scheme.
To validate the robustness of proposed multi-modal separation model in practical scenarios, the system was evaluated under the condition that one of the modalities is temporarily missing, invalid or corrupted. Experiments are conducted on a large-scale audio-visual dataset collected from YouTube (to be released) that spatialized by simulated room impulse responses (RIRs).
Experiment results illustrate that our proposed multi-modal framework significantly outperforms single-modal and bi-modal speech separation approaches, while can still support real-time processing.  \footnote{Separated samples are presented at https://moplast.github.io/}

\end{abstract}


%
\IEEEpeerreviewmaketitle

\section{Introduction}
\IEEEPARstart{T}{arget}
speech separation is to extract the speech of interest from an observed speech mixture \cite{Cherry1960Contribution}. In the speech processing literature, target speech separation has attracted tremendous interests for decades \cite{du2014speech}. With the entry into the deep learning era, most existing supervised approaches are based on spectrogram masking \cite{wang2014training,wang2018supervised,hershey2016deep,yu2017permutation,wang2018alternative}, where the weight (mask) of the target speaker at each time-frequency (T-F) bin of the mixture spectrogram is estimated. As a result, the multiplicative product between the mixture spectrogram and the predicted mask serves as the target speech spectrogram. However, these approaches only use audio information, termed audio-only approaches, often suffering from intense interferences in complex acoustic environment, such as noise and reverberation.

Recently, incorporating visual information into the speech separation system becomes an emerging research direction to improve the robustness and separation accuracy \cite{gabbay2018seeing,ephrat2018looking,afouras2018conversation,afouras2019my}. The principle is mainly twofolds: 1) The visual information (e.g., lip movements, face embeddings) is usually not affected by the acoustic environment;
2) It has been proved that the visual information is able to provide additional speech and speaker related cues. For example, speech content can be interpreted from the lip movements \cite{chung2017lip,chung2016lip}, which helps to improve the speech reconstruction quality \cite{ephrat2017improved}. Moreover, the face indicates the speaker identity information \cite{schroff2015facenet}. Besides the visual information, the feature representation vector of the speaker, termed speaker embedding, has also proved effective for extracting the target speaker's speech from the mixture signal \cite{wang2018deep,vzmolikova2017learning, wang2018voicefilter, vzmolikova2019speakerbeam}. Therefore, it is a promising direction to leverage the correlation and complementarity between different sorts of target speaker information for enhancing the performance of target speech separation.

Majority of previous multi-modal methods are established for monaural speech separation \cite{gabbay2018seeing,ephrat2018looking,afouras2018conversation,afouras2019my,wu2019time} and achieve state-of-the-art results on close-talk audio-visual speech separation datasets. In this work, aimed at enhancing the robustness and separation accuracy for far-field target speech separation, we present a general multi-modal framework.
The framework integrates multi-modal separation cues that extracted from the multi-channel speech mixture, the target speaker's lip movements and enrollment utterance. The idea is that, the acoustic target information can be blurry in the challenging acoustic environment, while the other modalities can provide complementary and steady information to increase the robustness. Also, we investigate on efficient multi-modality aggregation methods under this framework. A factorized attention-based aggregation method is proposed for fusing the high-level semantic information of multi-modalities at embedding level.
Finally, we address the modality robustness problem when one of the modalities is temporally noisy or unavailable.

In summary, this work makes three main contributions:
1) We introduce a multi-modal target speech separation framework, fully exploiting the target information, including directional information, lip movements and voice characteristics. To the best of our knowledge, this work is the first to integrate multi-modalities for far-field target speech separation;
2) Under the proposed framework, we investigate and propose several multi-modality fusion methods for target speech separation task;
3) Experiments demonstrate the robustness of proposed framework to the possible interferences from modality absence or noise.

\section{Related Works}
\label{sec:related_work}

In this section, we review related works in two areas: audio-only speech separation and audio-visual speech separation.

\subsection{Audio-only speech separation}
\label{subsec:audio_only}
Audio-only speech separation is extremely challenging under the single-microphone speaker-independent scenario, where no prior speaker information is available during evaluation. Majority of audio-only methods are based on spectrogram masking. 
Deep clustering \cite{hershey2016deep} first proposes to combine neural networks with the spectral clustering algorithm. Yu et al. \cite{yu2017permutation} designs a permutation invariant loss to reasonably assign the estimated mask to the reference speech during training. Lately, Luo et al. \cite{luo2019convtasnet} proposes fully convolutional time domain audio separation network (Conv-TasNet) to separate the speech mixture in time domain. It avoids phase reconstruction problem in spectrogram masking based methods and achieves state-of-the-art performance. When a multi-channel speech signal is available, microphone array based signal processing techniques can be leveraged to further enhance the separation performance. Well established spatial features, e.g., inter-channel phase difference (IPD), have been proven especially useful when combined at the input level for spectrogram masking based methods \cite{chen2018multi, wang2018multi, lianwu2019multi}.

Moreover, elaborately designed directional features that indicate the directional source's dominance in each T-F bin further improve the separation performance \cite{wang2019combining, gu2019neural}. Also, the separated speech can be associated with its corresponding directional feature, which enables target speech separation. However, these spatial cues extracted from the multi-channel signal suffer from the spatial ambiguity issue. The spatial ambiguity issue occurs when simultaneous speech come from close directions \cite{lianwu2019multi}, which makes the directional features less discriminative. In this case, if the target speaker separation network is only conditioned on directional information, it becomes uncertain about which speaker needs to be separated.

Apart from the directional information, target speech separation can also benefit from the prior knowledge of the speakers \cite{wang2018voicefilter,vzmolikova2017learning,wang2018deep}. The speaker embedding represents the speaker's voice characteristics and is usually extracted from an enrollment audio clip with a pre-trained neural network. With the aid of the speaker embedding (or speaker one-hot vector, speaker posterior in \cite{zmolikova2017speaker}), the separation network learns to extract and follow the target speaker over different frames. Furthermore, in \cite{xiao2019single}, in addition to speaker embedding of the target speaker, those of possible interfering speakers are also utilized to prompt the discrimination between speakers. But these methods have been only proven effective in close-talk corpora.

\subsection{Audio-visual target speech separation}
\label{subsec:a_v}

Multi-sensory integration using neural networks for acoustic scene perception have gained increasing interest in recent years. The studied areas include speech recognition \cite{afouras2018deep}, lip reading (predicting speech from silent video) \cite{chung2017lip}, acoustic event detection and localization \cite{owens2018audio}. In the same way, the audio-visual speech separation task and lip reading are closely linked. Gabbay et al. \cite{gabbay2018seeing} explores the correlation between the speaker's lip movements and speech spectrogram and proposes a video-to-sound method. However, it's a speaker-dependent approach since the video-to-sound model is separately trained for each speaker. Also, it is purely visually driven and has not employed the speech mixture signal. Then, Afouras et al. \cite{ephrat2018looking} introduces a large-scale audio-visual English dataset AVSpeech for training speaker-independent models. In \cite{ephrat2018looking}, the authors propose to jointly model the acoustic and visual components by making use of the speech mixture and the speaker's face embedding. Also, complex masks are served as the separation target for improving the phase reconstruction.  \cite{afouras2018conversation} shares the similar idea and designs an audio-visual framework, in which lip movements are served as visual information. These two approaches generalize well in real-world samples and unseen languages with consistent video and audio input. Recently, Afouras et al. \cite{afouras2019my} addresses the video obstruction problem when a speaker's lip is occluded by e.g. a microphone. To solve this problem, \cite{afouras2019my} combines the use of visual input and the speaker embedding of the target speaker. Therefore, when the speaker's mouth is occluded, voice characteristics of the target speaker can be relied on to compensate the target information. This approach is robust to partial video occlusions, hence a promising approach in practical applications. Wu et al. \cite{wu2019time} develops a time-domain audio-visual speech separation system, where short time Fourier transform (STFT) and inverse STFT (iSTFT) is replaced with a linear encoder and decoder. Therefore, the encoded audio representation is formulated in the real-value domain and complex phase estimation problem is avoided.

\section{Multi-Modal Multi-channel Separation}
\label{sec:proposed_system}

\begin{figure}[b]
    \centering
    \includegraphics[width=\linewidth]{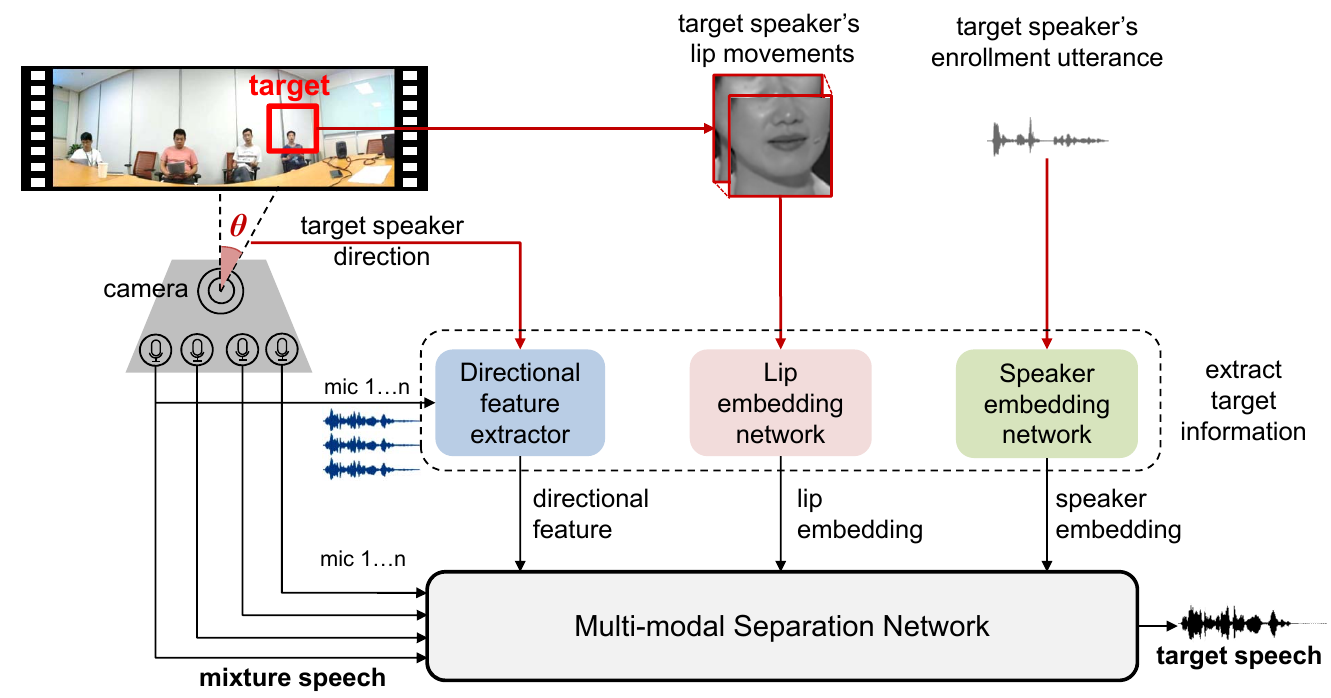}
    \caption{The diagram of proposed multi-modal target speech separation framework.
    }
    \label{fig:diagram}
\end{figure}

\begin{figure*}[b]
    \centering
    \includegraphics[width=\linewidth]{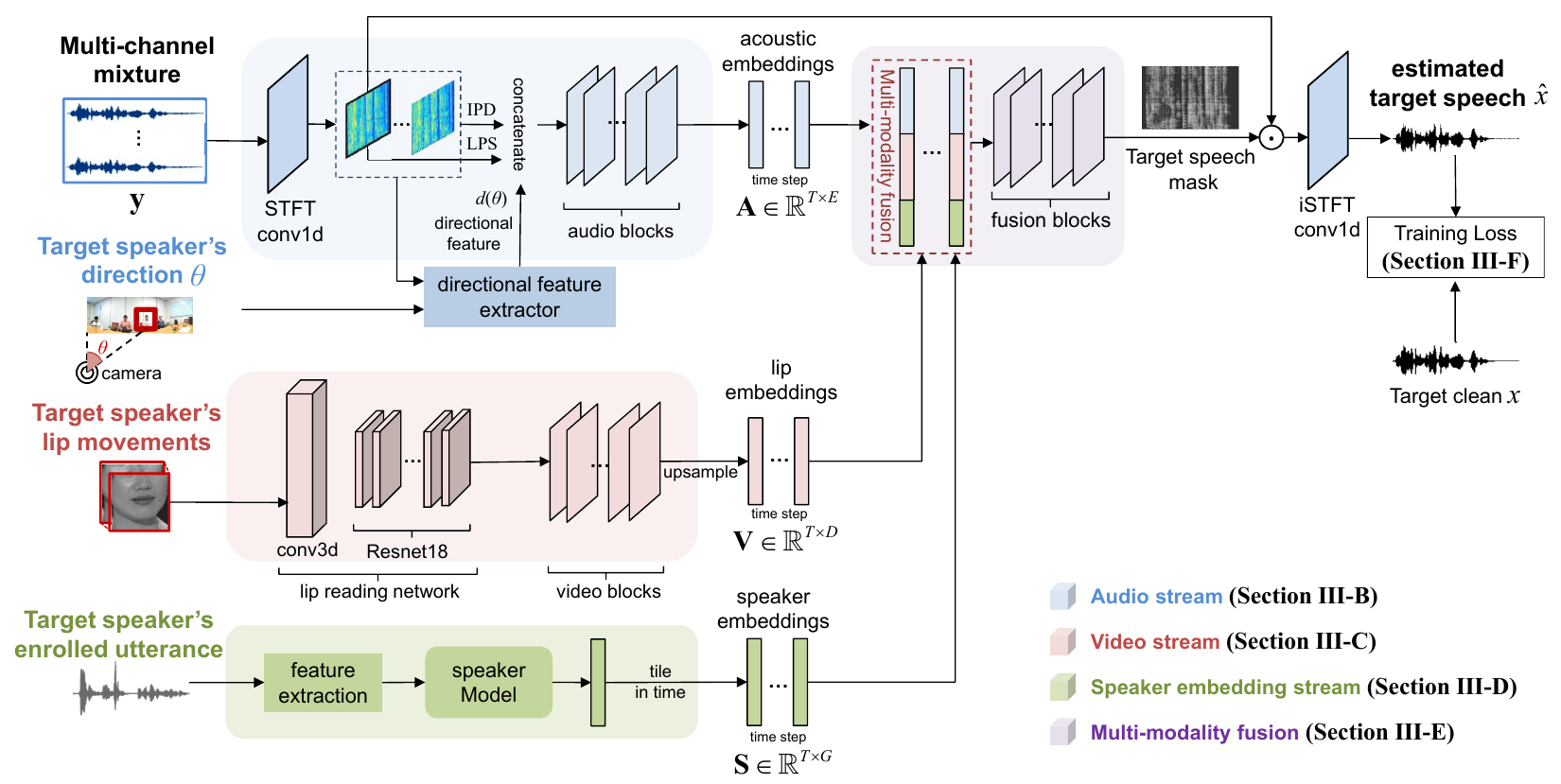}
    \caption{Our proposed multi-modal target speech separation framework with three streams: audio, video and speaker embedding stream. The video steam extracts frame-level lip embeddings from the lip video. The speaker embedding stream processes the enrollment audio(s) of the target speaker and generates an utterance-level speaker embedding. The audio stream takes the noisy multi-channel speech mixture and the target speaker's direction as input, extracting acoustic embeddings. Then, the multi-modality fusion module will combine these embeddings and feed into proceeding fusion blocks, which outputs a T-F mask for the target speaker. The output of our framework is the estimated target speech waveform.
    }
    \label{fig:proposed_system}
\end{figure*}

\subsection{Overview}

In this work, we address the task of separating the target speaker from a multi-channel speech mixture, by making use of target information from the target speaker's direction, lip movements and speaker embedding.
Previous works \cite{gu2019neural,wang2019combining,afouras2019my,ephrat2018looking,afouras2018conversation,wang2018voicefilter} have proposed to leverage part of these target information to perform the separation. As discussed in Section \ref{sec:related_work}, each kind of target related information has benefits and limitations. The directional information is quite effective for separating spatially diffuse sources, however it becomes invalid or even noisy when speakers are closely located. Although the visual information is not affected by the complex acoustic environment, the  lack of visual access to the speaker's face (e.g., turning and obstructions) may cause potential target absence. The speaker embedding works especially well for separating speakers of opposite genders, however the discriminability of speaker embeddings needs to be ensured via pre-training on a large-scale dataset.

In this work, we integrate all the target information into one framework, in order to achieve more superior and robust separation performance under challenging scenarios. 
As illustrated in Figure \ref{fig:diagram}, the proposed system is a multi-stream architecture which takes four inputs: (i) noisy multi-channel mixture waveforms, (ii) target speaker's direction calculated by face detection, (iii) video frames of cropped lip regions, (iv) enrollment audio(s) of the target speaker. The system directly outputs estimated monaural target speech, while all other interfering signals are suppressed.

\subsection{Audio Stream}
\label{subsec:a}
The detailed paradigm of audio stream processing is illustrated as the top stream in Figure \ref{fig:proposed_system}. A STFT convolution 1d layer is used to map the multi-channel mixture waveforms to complex spectrograms. Based on the complex spectrograms, the single-channel spectral feature and multi-channel spatial feature are extracted. Apart from the target speaker independent spectral and spatial features, a directional feature is extracted according to the spatial direction of target speaker. All of the features are then concatenated and fed into the audio blocks, which consist of stacked dilated convolutional layers with exponentially growing dilation factors, following \cite{luo2019convtasnet}. This design supports a long reception field to capture more sufficient contextual information. The output of the audio blocks are the acoustic embeddings $\mathbf{A} \in \mathbb{R}^{T \times E}$, where $E$ is the output convolution channels of the conv1d layers. On the system output side, an iSTFT convolution 1d layer is used to convert the estimated target speaker complex spectrogram back to the waveform.
Next, we will give a detailed description to the acoustic features, including the spectral, spatial and directional features.

\subsubsection{spectral feature}
To obtain the spectral feature from the $U$-channel raw mixture waveform $\mathbf{y}$, a standard STFT module is used for spectrum analysis.
STFT transforms the signal to a complex domain that can be decomposed into magnitude and phase components. Given a window function $w$ with length $N$, the multi-channel complex spectrogram $\mathbf{Y}$ calculated by standard STFT is written as:

\begin{equation}
\begin{split}
\mathbf{y}[n] \xrightarrow[]{\tt STFT} \mathbf{Y}_{t,f}
&=\overset{N-1}{\underset{n=0}{\sum}}\mathbf{y}[n]w[n-t]\exp{\left(-i\frac{2 \pi n}{N}f\right)}
\end{split}
    \label{eq:STFT}
\end{equation}

The logarithm power spectrum (LPS) of the reference channel (the first channel in this work) is served as the spectral feature, calculated by $\text{LPS} = \log(|\mathbf{Y}^1|^2) \in \mathbb{R}^{T\times F}$, where $\mathbf{Y}^1$ is the first channel of multi-channel complex spectrograms, $T$ and $F$ is the total frames and frequency bands of the complex spectrogram, respectively. In our implementation, the STFT operation is reformulated as a convolution kernel to enable on-the-fly computation \cite{wang2018end, wichern2018phase, gu2019neural} and speech up the separation process.

\subsubsection{spatial features}

As discussed in Section \ref{subsec:audio_only}, well-established spatial cues like IPDs have shown greatly beneficial for spectrogram masking based multi-channel speech separation methods \cite{lianwu2019multi,wang2018multi,chen2018efficient,wang2018spatial}. The standard IPD is computed by the phase difference between channels of complex spectrogram as:
\begin{equation}
\text{IPD}^{(m)}_{t,f}=\angle\mathbf{Y}^{m_1}_{t,f}-\angle\mathbf{Y}^{m_2}_{t,f}
\label{eq:ipd_ori}
\end{equation}
where $m_1$ and $m_2$ are two microphones of the $m$-th microphone pair, $M$ is the number of selected microphone pairs. Note that in our experiments, we don't have to use all pairs of microphones. To reduce the dimension of spatial features, we select $M$ microphone pairs with different spacings. $M$ pairs of $m:\{m_1, m_2\}$ are concatenated to form the IPD features: $\text{IPD}=[ ..., \text{IPD}^{(m)}_{t,f}, ..., ]_{M \times T \times F}$. The IPD extracts spatial information of all speakers in the mixture, so that we refer it as speaker-independent spatial feature.

\subsubsection{directional feature}
Given the direction of the target speaker, target-dependent directional feature can be extracted to provide explicit target information.
A location-guided directional feature (DF) for speech separation is introduced in \cite{chen2018multi}. The design principle lies in that if the T-F bin $(t,f)$ is dominated by the source from $\theta$, then $d_{t,f}(\theta)$ will be close to 1, otherwise close to 0. The DF is formed according to the direction of the target speaker, which measures the cosine distance between the steering vector and IPD:
\begin{equation}
\begin{aligned}
\centering
d_{t,f}(\theta) &=\overset{M}{\underset{m=1}{\sum}}
{\left < \mathbf{e}^{ \text{TPD}^{(m)}_{f} (\theta_t)},
\mathbf{e}^{\text{IPD}^{(m)}_{t,f}} \right>
}
\\
\text{TPD}_f^{(m)}(\theta_t) &= 2\pi f \varDelta_m \cos{\theta_t} /(f_{s}c) 
\end{aligned}
\label{eq:DF}
\end{equation}
where vector $\mathbf{e}^{(\cdot)}=\begin{bmatrix} \cos(\cdot) \\ \sin(\cdot) \end{bmatrix}$, $\text{TPD}^{(m)}_{f} (\theta_t) $ (Target-dependent Phase Difference) is the phase delay of a plane wave (with frequency $f$) experienced, evaluated at the $m$-th pair of microphones, travelling from angle $\theta_t$ (target speaker's direction at time $t$), $\varDelta_m$ is the distance between the $m$-th microphone pair, $c$ is the sound velocity and $f_s$ is the sampling rate.
In Eq. \ref{eq:DF}, we assume that all the speakers do not change their locations during speaking, i.e., $\theta_t=\theta$.
The pre-masking step in \cite{chen2018multi} is also applied to the DF to increase the discriminativity between speakers. Note that Eq. \ref{eq:DF} is reformulated so that it can be applied to general microphone array topology rather than the special seven-element microphone array used in \cite{chen2018multi}.

\textbf{How to obtain the target speaker's direction $\theta$. }
During training, the direction of target speaker $\theta$ is known, because the multi-channel audios for training are generated by simulation (see Algorithm \ref{alg:mm}). In practice, the direction of target speaker $\theta$ can be estimated by a face detection and tracking system \cite{dlib09}. Alternatively, audio-based localization methods can also be used to estimate the directions of multiple sound sources (with less than 10 degrees of mean absolute error \cite{chakrabarty2019multi}). However it remains uncertain that which direction of the sound corresponds to which speaker.
To address this issue, an additional speaker recognition system is required. The drawbacks of introducing this system are 1) an extra enrollment process is required; 2) comparing to the performance of the face recognition system \cite{dlib09}, the performance of state-of-the-art speaker verification systems is still far behind \cite{zhang2016end}.
Thus, for real-recorded samples, we use the face detection method to identify and track the target speaker in the video and estimate his/her direction based on the camera position. Since visual information is not affected by the acoustic environment, face detection based speaker localization method is more robust for our task. The details of face detection, recognition, tracking and speaker diarization are beyond the scope of this paper.  

\subsection{Video Stream}
\label{subsec:v}

For the video stream, the majority of previous audio-visual speech separation approaches \cite{afouras2018conversation, ephrat2018looking, afouras2019my, wu2019time} adopt the pre-training strategy. Before jointly training with the audio stream, they firstly set a lip reading objective to train the video stream, called lip reading network. The input of the lip reading network can either be a sequence of images of cropped lip regions \cite{afouras2018conversation} or the face embedding of the target speaker \cite{ephrat2018looking}. The network is trained to estimate the word-level or phone-level posteriors \cite{chung2017lip, chung2016lip}. The supervision information is formed with the speech transcription.

In this work, we try to separate the speech of Mandarin speakers. And due to the concern that there are a few lipreading datasets for Mandarin to train our lip reading network, we investigate the effects of joint training of both video and audio stream from scratch, only using the speech separation objective function (see Section \ref{subsec:e2e_training}).
As shown in Figure \ref{fig:proposed_system} (the middle stream), we follow the work in \cite{wu2019time, afouras2018conversation} and take gray frames as the input to the lip reading network. The structure of our lip reading network is similar to the one proposed by \cite{afouras2018conversation}, which consists of a spatio-temporal convolution layer and a 18-layer ResNet \cite{he2016deep}, to capture the spatio-temporal dynamics of the lip movements. The lip reading network is followed by several video blocks, each contains several dilated temporal convolutional layers with residual connections. ReLU and batch normalization \cite{ioffe2015batch} are also included in each block. The output of the video blocks are lip embeddings $\hat{\mathbf{V}} \in \mathbb{R}^{K \times D}$, where $K$ is the number of video frames and $D$ is the dimension of lip embedding. Since the time resolution of video and audio stream is different, we upsample the lip embeddings $\mathbf{V} \in \mathbb{R}^{T \times D}$, to synchronize the audio and video stream by nearest neighbor interpolation. The interpolated value at a query point is the value at the nearest sample point.

Since the supervision information is formed from audio domain, the video stream is propelled to discover the cross-domain correlations between the target speech and lip movements. One evident correlation is between the opening/closing of the mouth and voice activity. When a person's mouth is continuously open, there is a strong likelihood that he/she is speaking. Another less evident correlation is between the specific pattern of mouth movements and the phone. Since there is no supervision information for the lipreading objective, the learned lip embeddings may not discriminate all the phones well enough. However, the network may have the potential to learn phone clusters with distinct inter-differences.

\begin{figure}
	\centering
	\includegraphics[width=7cm]{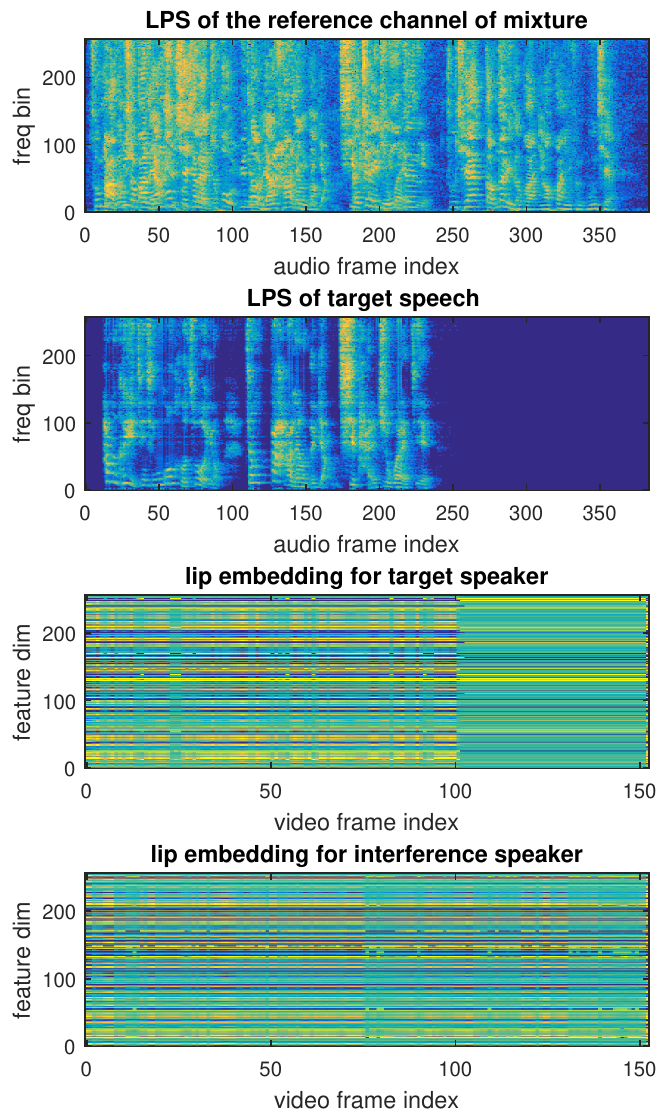}
	\caption{The illustration of lip embeddings of the target speaker and the interfering speaker when applied to a sample of the Mandarin dataset. The audio is analyzed with 32ms window and 16ms hop size. The frame rate of video is 25 frames per second (fps). }
	\label{fig:lip_emb}
\end{figure}

To intuitively observe the learned patterns of lip embeddings through joint training of the audio and video stream, Figure \ref{fig:lip_emb} visualizes the lip embeddings obtained from a sample of the Mandarin-mix dataset. Compared the LPS and the extracted lip embeddings of the target speaker, it is obvious that the beginning-ending points of speech contents in continuous speech can be inferred from the lip embeddings. Also, the lip embeddings of the target speech and those of the interfering speech exhibit different selection and emphasis on the embedding dimension.

\begin{figure}
    \centering
    \includegraphics[width=\linewidth]{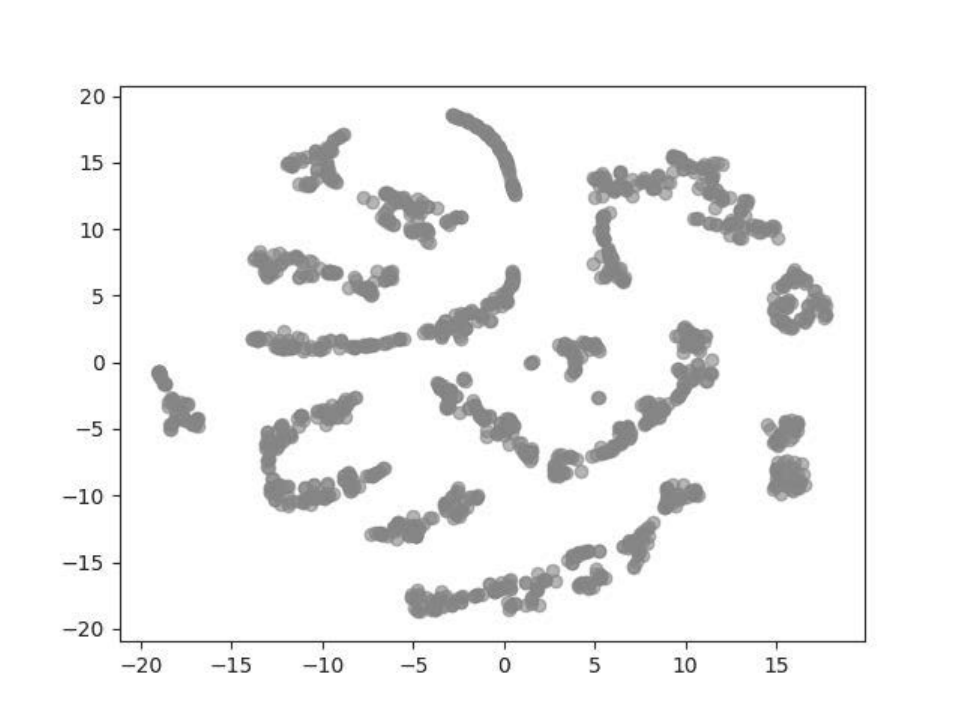}
    \caption{The t-SNE visualization of lip embeddings collected from 50 target lip videos. Each dot represents a video frame. The clusters imply the phonetic information has been learned in the lip embeddings.}
    \label{fig:tsne}
\end{figure}

Furthermore, Figure \ref{fig:tsne} visualizes the t-distributed stochastic neighbor embedding (t-SNE) of lip embeddings that collected from 40 lip videos. It is obvious that these lip embeddings naturally form clusters, which indicates the existence of mutual information cross video frames.

\subsection{Speaker Embedding}
\label{subsec:spk_emb}
As discussed in Section \ref{subsec:audio_only}, speaker embedding is a kind of bias signal that informs the separation network of the target information and enables target speaker separation. Here, we introduce a pre-trained speaker model and utilize its produced embedding to characterize the target speaker.
The speaker model was pre-trained on speaker verification task \cite{zhang2018text}, with 4 convolution layers followed by a fully connected layer. To achieve more discriminative speaker embeddings, self-attention is adopted as the frame-level feature aggregation strategy.
The input to the speaker model is an enrollment utterance of the target speaker. The speaker model outputs the utterance-level speaker embedding $\mathbf{s} \in \mathbb{R}^{1\times G}$, where $G$ is the speaker embedding dimension. To match the time steps of the audio stream, the speaker embedding is tiled in time as $\mathbf{S}=[...,\mathbf{s}_t,...]\in \mathbb{R}^{T\times G}$, where $\mathbf{s}_t=\mathbf{s}$.

\subsection{Multi-modality Fusion}
\label{subsec:aggregation}
As described in above sections, three kinds of target information are derived from a set of media sources, including acoustic embeddings from multi-channel speech, lip embeddings from the video and speaker embedding from the target speaker's enrollment utterance.
In order to learn effective target speech extraction from multi-modal information, in this section, we will describe and discuss the investigated methods on fusing these modalities.

\subsubsection{Concatenation}
The most common approach to integrate the multi-modal embeddings is to simply concatenate them along the feature axis. This fusion method has been widely used in previous audio-visual speech separation works \cite{ephrat2018looking,afouras2018conversation,afouras2019my}. The subsequent network is expected to automatically learn the interaction between cross-domain embeddings. In this way, all the modalities are treated equally and the potential correlation between modalities may not be effectively explored.

\begin{figure}[b]
    \centering
    \includegraphics[width=4.5cm]{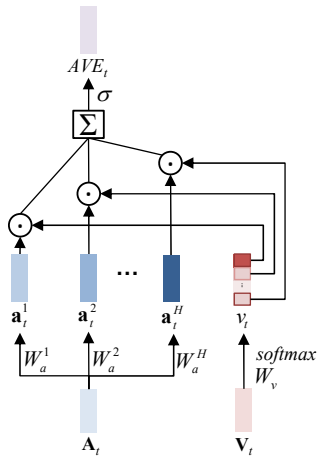}
    \caption{The illustration of factorized attention when fusing audio embeddings with video embeddings.}
    \label{fig:factor}
\end{figure}

\subsubsection{Factorized Attention}

In recent speech recognition work, a factorized layer is proposed \cite{delcroix2015context} for fast adaptation to the acoustic context. In speech recognition literature, a factor characterizes a set of speakers or a specific acoustic environment \cite{yu2012factorized}. The factorized layer uses a set of parameters to process each acoustic class and these parameters are dependent on external factors that represent the acoustic conditions.

Inspired by this, we propose to factorize the acoustic embeddings into a set of acoustic subspaces (e.g., phone subspaces, speaker subspaces) and utilize information from other modalities to aggregate them with selective attention. The other modalities can also provide information related to the acoustic condition, such as voice activity interpreted from the opening and closing of the mouth, and voice characteristics contained in the speaker embedding.

Specifically, we take the audio-visual fusion as an example, illustrated in Figure \ref{fig:factor}.
Firstly, the acoustic embeddings $\mathbf{A}$ are factorized into different acoustic subspaces with parallel linear transformations ${W}_a^1,{W}_a^2,...,{W}_a^H$, where $H$ is the number of subspaces and the acoustic representation in $h$-th subspace at the $t$-th time step is denoted as ${\mathbf{a}_t^h}=\mathbf{A}_t W_a^{h} \in \mathbb{R}^{1\times P}$, where $P$ is dimension of each subspace.
Then, the lip embeddings $\mathbf{V}$ are mapped from the $D$-dimensional space to a $H$-dimensional space, where each dimension $h$ is expected to contain bias information that corresponds to the $h$-th acoustic subspace. Next, these mapped lip embeddings are passed to a softmax layer and then produce the estimated posterior for each subspace at each time step, calculated as $\mathbf{v}=\text{softmax}(\mathbf{V}W_v)=[v^1,{v}^2,...,{v}^H]\in \mathbb{R}^{T\times H}$. Finally, the fused audio-visual embedding (AVE) is obtained by summing up the weighted contribution of different acoustic subspaces:
\begin{equation}
\textit{AVE}_t=\sigma \left(\sum_{h=1}^{H} v^h_t \mathbf{a}^h_t \right)
\label{eq:factor}
\end{equation}
where $\sigma$ is the sigmoid activation function.
As for using factorized attention for acoustic and speaker embedding fusion, the audio-speaker embedding can be calculated by
$\textit{ASE}_t=\sigma{ \left( \sum_{h=1}^{H} \textit{softmax}(\mathbf{s}_{t}W_s)^h \mathbf{a}^h_t\right )}$, where $W_s$ is the weight matrix that converts the speaker embedding $\mathbf{s}_t$ from the speaker space to acoustic subspaces.

Compared to direct concatenation, the factorized attention sums over all possible speakers or acoustic context guided by cross-modal information. The interaction of embeddings of different modalities in various subspaces enables the deep semantic information capturing and selection.

\subsubsection{Rule-based Attention}
The motivation for fusing multi-modalities with attention lies in that, the effectiveness and significance of each modality depends on the case. For example, when the speakers come from close directions, the discriminability of spatial and directional features may be weaker.
In general, our strategy is to foster strengths and circumvent weaknesses among features of different modalities. Therefore, the network should selectively attend to discriminative modalities and ignore the other ones. Following our previous work \cite{gu2019neural}, we compute the attention using the priori knowledge of angle difference between speakers. Specifically, when the angle difference $ad$ between speakers is small, the weight score that applied to spatial and directional features is relatively low, calculated as:
\begin{equation}
att(ad)=2*\max
\left (\sigma(ad)-0.5,0 \right)
\label{eq5}
\end{equation}
where $\sigma(ad)=1/(1+exp(-w(ad-b)))$ is the sigmoid score denotes how much emphasis should be put on spatial features and directional feature, $w$ and $b$ are trainable parameters.
Note that the rules can take other factors into consideration, such as the whether the face is sufficiently frontal-facing, etc.

\subsubsection{Fusion of three modalities}
To reduce the learning difficulty, for fusion of three modalities, we adopt a hierarchical fusion strategy. Specifically, the three-modality fusion is divided into two stages, proceeding from unimodal to bimodal embeddings and then bimodal to trimodal embeddings \cite{majumder2018multimodal}. Also, different fusion methods can be adopted at each stage. For example, the acoustic and speaker embeddings are firstly fused using the factorized attention method. Then, the fused ASE is concatenated to the lip embeddings and combined into the trimodal embeddings. The details will be described in Section \ref{sec:exp}.

\subsection{End-to-End Training}
\label{subsec:e2e_training}

The fusion blocks are followed by a $1\times1\textit{-conv}$ layer and a nonlinear activation function (rectified linear Unit (ReLU) in this work), which produces the estimated magnitude mask $\in \mathbb{R} ^{T\times F}$ for target speech. Then, the estimated target speech complex spectrogram can be obtained by multiplying the reference channel of mixture complex spectrogram $\mathbf{Y}^1$ by the estimated mask. Finally, the iSTFT operation is used to convert the estimated target speech spectrogram back to the waveform.

To optimize the network from end to end, instead of using a time-domain mean squared error (MSE) loss, the speech separation metric scale-invariant signal-to-distortion (SI-SDR) is used to directly optimize the separation performance, since it has been proven better for speech separation \cite{bahmaninezhad2019comprehensive}. The SI-SDR is defined as:
\begin{equation}
\left\{
\begin{array}{lr}
x_{\text{target}}:=\frac
{\left<\hat{x}, x\right>x}
{\left\|x\right\|_{2}^{2}} \\
e_{\text{noise}}:=\hat{x}-x_{\text{target}}  \\
\text{SI-SDR}:=10\log_{10}\frac
{\left\|x_{\text{target}}\right\|_{2}^{2}}
{\left\|e_{\text{noise}}\right\|_{2}^{2}}
\end{array}
\right.
\end{equation}
where $x$ and $\hat{x}$ are the reverberant clean and estimated target speech waveform, respectively. The zero-mean normalization is applied to $x$ and $\hat{x}$ to guarantee the scale invariance.

\section{Experiments procedures}

\subsection{Dataset}

The audio-visual corpus used for experiments is collected from Youtube, in which Mandarin accounts for the vast majority.
To select relatively high quality videos, a signal-to-noise (SNR) estimator is used to filter out videos with low SNR speech, and a face detection model is used to further remove the videos without the speaker face. After selection, there are about 1,000 speakers and 53,000 clean utterances in total.
A mouth region detection program is run on the target speaker's video to capture the the lip movements. The sampling rate for audio and video are 16 kHz and 25 fps respectively.

\begin{figure}[b]
    \centering
    \includegraphics[width=\linewidth]{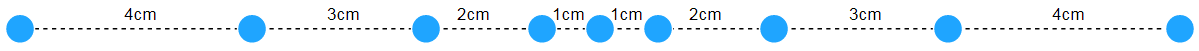}
    \caption{The 9 element non-uniform linear array layout}
    \label{fig:arrayLayout}
\end{figure}

The multi-talker multi-channel mixtures are simulated with steps in Algorithm \ref{alg:mm}. The simulated dataset contains 160,000, 15,000 and 1,200 multi-channel noisy and reverberant mixtures for training, validation and testing. The speakers in the training set and test set are not overlapped, which means our approach is evaluated under speaker-independent scenario. The duration of each utterance is ranging from 1.0 to 15 seconds and the average duration is about 4.5s. We use a 9-element non-uniform linear array, with spacing 4-3-2-1-1-2-3-4 cm, as shown in Figure \ref{fig:arrayLayout}. The multi-channel audio signals are generated by convolving single-channel signals with Room Impulse Responses (RIRs) simulated by image-source method \cite{ISM}.
The room size is ranging from 4m-4m-2.5m to 10m-8m-6m (length-width-height). The speakers and the microphone array randomly located in the room at least 0.3m away from the wall. The distance between the speaker and microphones ranges from 1m to 5m. The reverberation time T60 is sampled in a range of 0.05s to 0.7s. The signal-to-interference rate (SIR) is ranging from -6 to 6 dB. Also, noise with 18-30 dB SNR is added to all the multi-channel speech mixtures.

To evaluate system performance of both non-overlapped and overlapped speech, we consider three scenarios for the synthetic examples generation: 1 speaker, 2 speakers and 3 speakers, respectively accounts for 49\%, 30\% and 21\% in the test dataset. For the overlapped speech of 2 and 3 speakers cases, samples with angle difference of 0-15$\degree$, 15-45$\degree$, 45-90$\degree$ and 90-180$\degree$ respectively accounts for 16\%, 19\%, 11\% and 5\% in the test dataset, where the angle difference is defined as the smallest degree difference between the target speaker and other interfering speakers.

\begin{algorithm*}[t]
	\caption{Data Simulation Process of Audio-visual Mandarin Dataset.}
	\label{alg:mm}
	\DontPrintSemicolon
	\SetAlgoNoLine
	\KwIn{audio-visual Mandarin corpus}
	\KwOut{audio-visual spatialized noisy and reverberant Mandarin-mix}
	\For {1: total mixture number}
	{
	1) Sample the number of speakers $C$ in the mixture from [1, 2, 3];\\
	2) Randomly select $C$ videos from the audio-visual Mandarin corpus;\\
	3) Run face detection \cite{dlib09} on each video and capture the corresponding lip movements;\\
	4) Extract utterance-level speaker embeddings using the enrolled utterances of target speakers;\\
	5) Sample mixed SIR uniformly from [-6:6] dB for each video's audio stream;\\
	6) Sample room size [$r_x$, $r_y$, $r_z$] from 4m-4m-2.5m to 10m-8m-6m;\\
	7) Sample T60 of room from [0.05, 0.7] seconds;\\
	8) Generate microphone array position in the room randomly. The array is at least 0.3m away from the wall;\\
	9) Generate speakers position in the room randomly. The distance between speakers and array is [1, 5] m;\\
	10) Sample noise from a 20-hours data set including music, TV, office, kitchen, babble etc. noises;\\
	11) Generate impulse responses using RIR generator;\\
	12) Convolve each single-channel source with corresponding RIR to generated reverberated multi-channel source;\\
	13) Scale reverberated sources with sampled SIR;\\
	14) Add these scaled and reverberated sources along with selected noise under [18:30] SNR to obtain the final mixture. Note each mixture is associated with the target speaker's position (direction), lip movements and speaker embeddings. The length of final simulated utterance is decided by the longest utterances among target speech and interfering speech.
	}
\end{algorithm*}

The data will be released and more details will be described in \cite{ailab2019large}.

\subsection{Features}

\noindent\textbf{Audio}. For short time Fourier transform (STFT) setting, we use 32ms sqrt hann window and 16ms hop size. Therefore, the frame size and shift are 512 and 256 points, respectively. 512-point FFT is used to extract 257-dimensional LPS. The LPS is computed from the first channel waveform of speech mixture. IPDs are extracted between 5 microphone pairs (1, 9), (1, 5), (2, 5), (5, 7) and (5, 6). These pairs are selected considered that different spacings between microphones can be sampled. For calculating the DF, we use the same microphone pairs for TPDs. During both training and evaluation sessions, the ground truth target speaker's direction is used for computing the DF. The total dimension for acoustic features are 7$\times$257=1799. The dimension of acoustic embedding is $E$=256 in all experiments.

\noindent\textbf{Lip video}. Each input frame of the video is gray with the size of 112$\times$112$\times$1 (height$\times$width$\times$channel). The dimension of lip embeddings is the same in all experiments, i.e., $D$=256.

\noindent\textbf{Enrollment}. For each speaker, there are about 10 utterances for enrollment on the average (about 30-40 seconds). The overall speaker embedding is obtained by averaging all the utterance-level speaker embeddings. The dimension of speaker embedding is 128.

\subsection{Network structure}

\begin{figure}[t]
    \centering
    \includegraphics[width=5.5cm]{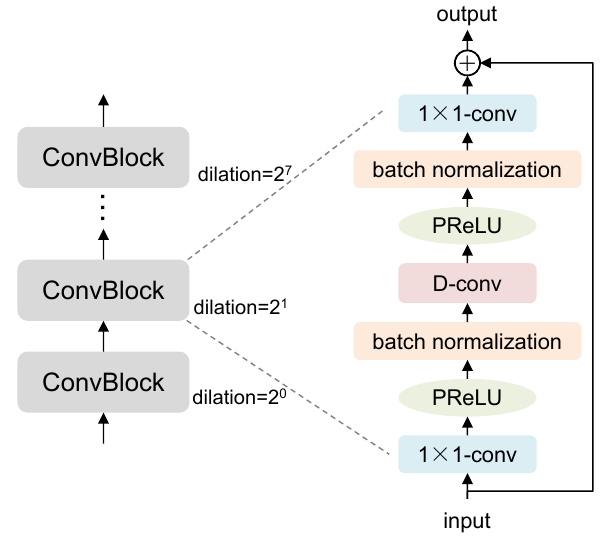}
    \caption{The illustration of convolutional blocks (ConvBlocks). Each ConvBlock consists of a 1$\times$1-conv layer, a depth-wise separable convolution layer ($D-conv$), with PReLU activation function and normalization added between each two convolution layers. Also, residual connection is added in each block.}
    \label{fig:convblock}
\end{figure}

\noindent\textbf{Audio processing}. After the concatenation of spectral and spatial features, they are fed into proceeding audio blocks. The design of these blocks followed the version 2 of \cite{luo2018tasnet}, as illustrated Figure \ref{fig:convblock}. The number of channels in $1\times1\textit{-conv}$ layer is set as 256. For the $D-conv$ layer, the kernel size is 3 with 512 channels. Batch normalization instead of global layer normalization is adopted considering the processing speed. Every 8 convolutional blocks are packed as a repeat, with exponentially increased dilation factors $2^0, 2^1, ..., 2^7$.

\noindent\textbf{Video processing}. The structure of the lipnet is the same as \cite{chung2017lip}. The extracted lip embeddings are then passed to video blocks, including 5 convolutional blocks. The block design is similar to that of audio blocks, including depth-wise separable convolution layer, ReLU and normalization and residual connection.

\noindent\textbf{Fusion methods}. For factorized attention, the factor number $H$ is set to 10 empirically. The dimension for acoustic embedding in each subspace is $\mathbb{R} ^ {T\times 256}$. As a result, the weight matrix $W^{h}_a \in \mathbb{R} ^{256\times 256}$ for each audio linear layer and $W_v \in \mathbb{R} ^{256\times 10}$ for the video linear layer. Also, a softmax layer followed the video linear layer to compute the posteriors of each subspace. For rule-based attention, according to \cite{gu2019neural}, the $w$ and $b$ is initialized with -0.5 and 10, respectively. After the fusion of multi-modalities, the fused embeddings are passed to fusion blocks. Fusion blocks include $N_f$ times of repeats, which contains $N_f \times 8$ convolutional blocks, in our experiments $N_f$ is set as 3, following \cite{wu2019time}. The number of convolution channels is 256.

\subsection{Training procedure}

The training of model includes two stages. First, the speaker model is pre-trained with the speaker verification on a Mandarin dataset first. Later, it is freezed and utilized to extract speaker embeddings from all the enrollment audios. Second, the audio and video streames are jointly trained from scratch. The multi-modal network is trained with utterance-level mixtures, using Adam optimizer with early stopping. Initial learning rate is set to 1e-3. If there is no improvement for consecutive 4 epochs on validation loss, the learning rate will be halved.

\subsection{Evaluation metrics}

Following the recent advances in speech separation metrics \cite{le2019sdr}, average SI-SDR is adopted as the main evaluation metric. Also, following the common practice, perceptual estimation of speech quality (PESQ), short-time objective intelligibility (STOI) and average SDR \cite{vincent2006performance} are also used to measure the speech quality. To further assess the intelligibility of the estimated speech, we use the Yitu automatic speech recognition (ASR) system to compute the speaker attributed word error rate (WER) \cite{galibert2013methodologies} between the separated speech and the ground truth target speech. Speaker attributed WER refers to the sum of transcription errors attributed to the target speaker divided by the reference words.
Since we do not perform speech dereverberation, we consider the reverberant clean speech as the reference for all the metric computation.

All the trained models are evaluated without knowing the number of sources in the mixtures, since the models perform target speech separation. Apart from the overall performance, we also evaluate the performances under different ranges of angle difference between speakers, and performances under different speaker mixing conditions. The relative performance difference varying from scenarios may help us give a more comprehensive assessment to the model.

\section{Results and Analysis}
\label{sec:exp}

\subsection{Fusion approaches}
\label{subsec:fusion}

In this subsection, we will investigate different multi-modality fusion approaches, including the fusion of audio and speaker embedding (audio-speaker), audio and video (audio-visual) and audio, video and speaker embedding (multi-modal). The baseline is set as DF-only model that only trained with spectral, spatial and directional features (LPS+IPDs+DF).

Table \ref{tab:as_fusion} compares the performance of the audio-speaker models using different fusion methods, including directly concatenation and factorized attention, trained with all data.
Both concatenation and factorized attention do not improve the overall performance, possibly due to the unsatisfactory discrimination between speaker embeddings. However, factorized attention boosts the performance from 7.1dB to 7.7dB under small angle different range. Since DF's discriminability significantly decreases under small angle difference case, speaker embedding may play an important role in providing the target-related information.

\begin{table}[t]
  \caption{SI-SDR (dB) performances of audio-speaker models adopting different fusion methods.}
  \label{tab:as_fusion}
  \centering
  \begin{tabular}{l|c|p{13 pt}<{\centering}p{23 pt}<{\centering}p{23 pt}<{\centering}p{13 pt}<{\centering}|c}
    \toprule
    \multirow{2}{*}{\textbf{Method}} &
    \multirow{2}{*}{\textbf{\#Param}} &
    \multicolumn{5}{c}{\textbf{SI-SDR  (dB)}} \\
    & & $<$15$\degree$ & 15-45$\degree$ & 45-90$\degree$ & $>$90$\degree$ & Ave. \\
    \hline
    DF-only & 9.6M & 7.1 & 9.3 & 10.5 & 10.8 & 9.1  \\
    \hline
 Concatenation  &9.7M &  7.2 & 9.3 & 10.5 & 10.8 & 9.1\\
 Factorized att. &10.3M & 7.7 & 9.0 & 10.5 & 10.6 & 9.1\\
    \bottomrule
  \end{tabular}
\end{table}

Table \ref{tab:av_fusion} compares the performance of the audio-visual models using different fusion methods. These models are trained only with overlapped data to save the training time.
Both directly concatenation and rule-based attention do not show clear performance gain over DF-only model. Among three audio-visual fusion methods, factorized attention exhibits the best overall performance, owing to the benefits brought by subspace factoring and the learnable attention.

\begin{table}[t]
  \caption{SI-SDR (dB) performances of audio-visual models adopting different fusion methods, trained on overlapped data.}
  \label{tab:av_fusion}
  \centering
  \begin{tabular}{l|c|p{13 pt}<{\centering}p{23 pt}<{\centering}p{23 pt}<{\centering}p{13 pt}<{\centering}|c}
    \toprule
    \multirow{2}{*}{\textbf{Method}} &
    \multirow{2}{*}{\textbf{\#Param}} &
    \multicolumn{5}{c}{\textbf{SI-SDR (dB)}} \\
    & & $<$15$\degree$ & 15-45$\degree$ & 45-90$\degree$ & $>$90$\degree$ & Ave. \\
    \hline
    DF-only &9.6M & 7.4 & 8.7 & 10.5 & 11.5 & 9.0 \\
    \hline
 Concatenation  &21.4M &  7.5 & 9.0 & 10.5 & 11.4 & 9.1\\
 Factorized att. &21.9M & 7.6 & 9.3 & 10.7 & 11.5 & 9.3\\
 Rule-based att. &21.4M & 6.3 & 9.4 & 10.5 & 11.9 & 8.9 \\
    \bottomrule
  \end{tabular}
\end{table}

\begin{table}[b]
  \caption{SI-SDR (dB) performances of multi-modal models adopting different fusion method combinations, trained on overlapped data. a-s and a-v represent the audio-speaker and audio-visual model, respectively.}
  \label{tab:avs_fusion}
  \centering
  \begin{tabular}{cc|p{23pt}<{\centering}|p{13pt}<{\centering}ccp{13pt}<{\centering}|c}
    \toprule
    \multicolumn{2}{c|}{\textbf{Fusion method}} &
    \multirow{2}{*}{\textbf{\#Param}}&
    \multicolumn{5}{c}{\textbf{SI-SDR (dB)}} \\
    a-s & a-v & & $<$15$\degree$ & 15-45$\degree$ & 45-90$\degree$ & $>$90$\degree$ & Ave. \\
    \hline
    \multicolumn{2}{c|}{DF-only} &9.6M & 7.4 & 8.7 & 10.5 & 11.5 & 9.0 \\ 
    \hline
 concat.  & concat. &21.4M &  7.1 & 9.1 & 10.6 & 11.4 & 9.0 \\
 fac. att.  & concat.&22.1M & 8.5 & 9.6 & 10.8 & 11.8 & 9.7 \\ 
 fac. att.  & fac. att.&22.7M & 8.2 & 9.4 & 10.8 & 11.7 & 9.5\\
    \bottomrule
  \end{tabular}
\end{table}

Table \ref{tab:avs_fusion} lists the performances of multi-modal models adopting different fusion methods, trained on the overlapped data. Specifically, the experimental setup of each multi-modal fusion method is as following:

\begin{itemize}
    \item \emph{concat. + concat.}: The fusion of acoustic, lip and speaker embeddings is performed after all the audio blocks. These embeddings are concatenated along the feature axis at each time step and assigned with equal weight. The fused embedding is interpreted as $\text{FE}=\textit{concat}(\mathbf{A},\mathbf{V},\mathbf{S})$. The fusion blocks consist of 3 repeats.

    \item \emph{fac. att. + concat. }: Firstly, the fusion of acoustic and speaker embedding is done using factorized attention method after all the audio blocks. Then, the fused embeddings are then concatenated with lip embeddings, written as $\text{FE}=\textit{concat}(ASE,\mathbf{V})$. Finally, these fused embeddings are passed to 3 repeats of fusion blocks.

    \item \emph{fac. att. + fac. att.}: The acoustic embeddings are firstly fused with speaker embedding after audio blocks using factorized attention. Then, the ASE is fed into proceeding 2 repeats of fusion blocks and more abstract and high-level embeddings are generated. Next, these embeddings are fused with lip embeddings by factorized attention. Finally, the fused multi-modal embeddings are further passed to a extra repeat of fusion blocks. Our intention to put off the fusion with lip embeddings lies in that, at deeper layers, the phonemic information may be better abstracted from the audio, which may make the fusion with lip embeddings more efficient.
    \end{itemize}{}

As shown in Table \ref{tab:avs_fusion}, the best result is presented by fusion method of factorized attention (audio-speaker) and concatenation (audio-visual). The factorized attention for both audio-speaker and audio-visual fusion is more effective than concatenation. However, it does not provide expected satisfactory performance, possibly due to the late fusion of lip embeddings and acoustic embeddings.

\begin{table*}[t]
  \caption{SI-SDR (dB), SDR (dB), PESQ, STOI, WER (\%) and RTF results of target separation models with different modalities.}
  \label{tab:feature}
  \centering
  \begin{tabular}{ccc|ccc|cccc|c|c|{c}|c|c|c}
    \toprule
    \multicolumn{3}{c|}{\textbf{Features}} &
    \multicolumn{8}{c|}{\textbf{SI-SDR (dB)}} &
    \multirow{2}{*}{\textbf{SDR (dB)}} &
    \multirow{2}{*}{\textbf{PESQ}} &
    \multirow{2}{*}{\textbf{STOI}} &
    \multirow{2}{*}{\textbf{WER (\%)}} &
    \multirow{2}{*}{\textbf{RTF}}\\
    DF & Lip & Spk  & 1spk &2spk &3spk & $<$15$\degree$ & 15-45$\degree$ & 45-90$\degree$ & $>$90$\degree$  & ave. &&&&\\
    \hline
    \multicolumn{3}{c|}{Mixture} &21.1 & -0.2 & -2.1 & -1.1 & -1.1 & -0.9 & -0.2 & 9.8 & 9.8 &2.77 & 0.83 & 48.8 & - \\
    \hline
    \checkmark  & &  & 24.4 & 10.2 & 7.6 & 7.2 & 9.3 & 10.5 & 10.8 & 16.5 & 16.9 &3.24 &0.91 & 11.3 & 0.0040 \\
    & \checkmark & & 24.0 & 9.5 & 7.0 & 7.7 & 7.7 & 8.3 & 8.8 & 16.2 & 16.6 & 3.01 & 0.89 &19.6 & 0.0080 \\
    & & \checkmark  & 23.4 & 6.9 & 3.7 & 5.7 & 4.8 & 6.4 & 6.5 & 14.4 & 14.8 &2.98 & 0.87 & 14.7 & 0.0039 \\
    \hline
    \checkmark & &  \checkmark  & 24.5 & 10.3 & 7.7 & 7.7 & 9.0 & 10.5 & 10.6 & 16.7 & 17.1 & 3.25 & 0.91 & 10.5 & 0.0049 \\
    \checkmark & \checkmark &  & 24.7 & 10.8 & 8.2 & 8.6 & 9.6 & 10.8 & 11.7 & 17.1 & 17.5 & 3.28 & 0.92 & 10.3 & 0.0089 \\
    \checkmark & \checkmark & \checkmark  & 24.8 &10.9 & 8.3 & 8.6 & 9.7 & 10.9 & 11.8 & 17.2 & 17.6  & 3.28 & 0.92 & 10.0 & 0.0091 \\
    \bottomrule
  \end{tabular}
\end{table*}

\subsection{Impact of different modalities}

After the investigations of modality fusion approaches, in this subsection, we further analyze the impact of different modalities. The aim is to verify each modality, along with the reasonable multi-modality fusion, is effective in multi-channel target speech separation.

Table \ref{tab:feature} reports the performances of target separation models with different modalities input. All the models included LPS and IPDs in input and trained on the whole training dataset. The fusion method for models with more than one modality is chosen according to the best result achieved in Section \ref{subsec:fusion}. Specifically, factorized attention for audio-speaker, factorized attention for audio-visual, and factorized attention + concatenation for multi-modal. Also, the real-time factor (RTF) is also reported for computation measurement. The real time factor is defined as the GPU processing time (s) divided by the audio time (s). The RTF result is evaluated on the whole test set and it indicates that whether the model is fast enough for real-time processing.

From Table \ref{tab:feature}, it's obvious that DF makes a significant contribution to the overall performance, compared to speaker embedding and lip information. Also, the computation complexity for DF-only model is relatively low, achieving a real-time factor of 0.4\% on the GPU. However, the performance of DF-only model under small angle difference is poorer than that of lip-only model.
With the aid from lip movement information or speaker embedding, both audio-speaker and audio-visual models have a relative improvement on overall performance, especially under small angle difference range. The multi-modal model exhibits the best performance: 3.7dB, 11.1dB, 10.4dB SI-SDR improvement under 1spk, 2spk, and 3spk case respectively. Also, the multi-modal model achieves the lowest WER (10\%) among all the models. This confirms the effectiveness of our proposed multi-modality exploitation and integration approach. Although an increased RTF is observed, the process can still be achieved in real-time (i.e., RTF $< 1$).

\begin{figure}[t]
    \centering
    \includegraphics[width=6.5cm]{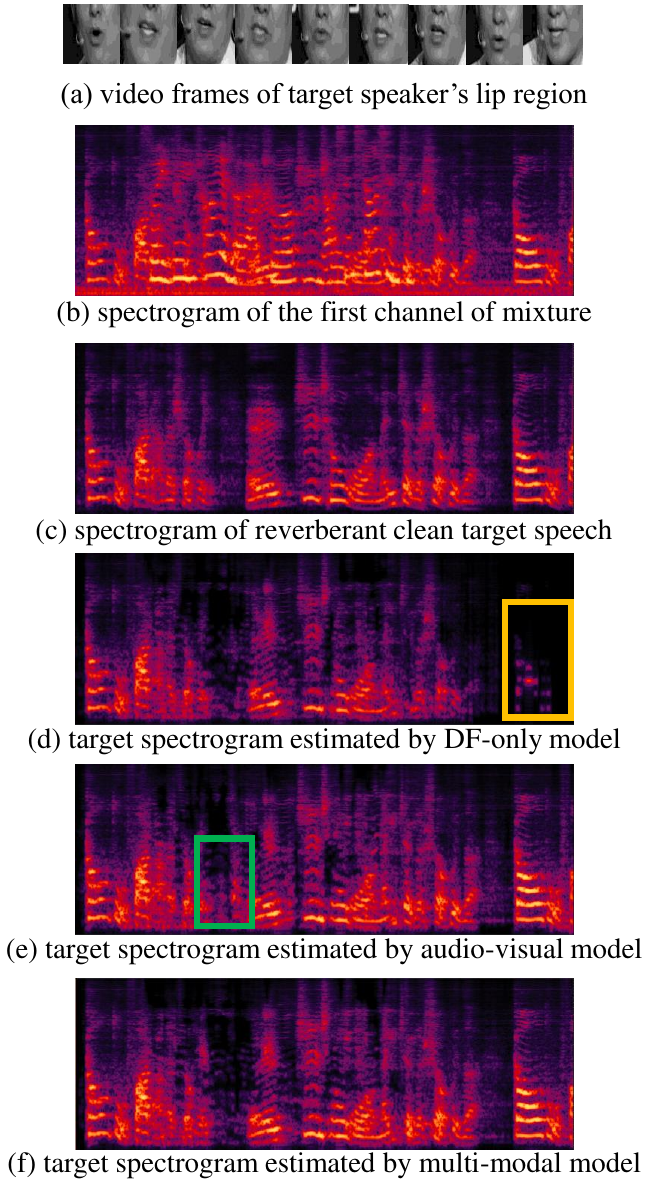}
    \caption{An example for target speech separation on the task of two-speaker separation. DF-only model (d) over-suppresses the target speech in the yellow box when the speaker slightly turns his face. Audio-visual model (e) may suffer from interference leakage when the speaker opens his mouth without speaking. }
    \label{fig:demo}
\end{figure}

In order to intuitively verify the benefits brought by multi-modal integration, Figure \ref{fig:demo} presents an example of separation results estimated by DF-only model, audio-visual model and multi-modal model, respectively. From Figure \ref{fig:demo}(d) we can see that the DF-only model loses the target speech in the yellow box. This may happen when the target speaker temporally turned his face, then the direction estimated by face detection may deviate from the ground truth. Also, the result estimated by audio-visual model (Figure \ref{fig:demo}(e)) did not filter out the interfering sound in the green box. This is probably due to that the target speaker opens his mouth while not actually speaking. With all the target information available, multi-modal model produces the best estimation result (Figure \ref{fig:demo}(f)), compared to the target speech spectrogram (Figure \ref{fig:demo}(c)).

\subsection{Modality Robustness}

There may be many cases in practical that one of the modalities is unavailable or unreliable. In order to demonstrate the robustness of our multi-modal model in real-world scenarios, we tested it under two particular cases: temporarily missing lip information and estimation error of the target speaker's direction.

\subsubsection{Impact of missing lip information}

In practical, the lip information may be invalid in many cases. For example, the transmission of high-resolution video may be not stable enough, thus the frames may drop randomly. Moreover, the target speaker may temporarily turn his face away from the camera, or his lip may be obstructed by the microphone. We regard these scenarios as the missing of lip information. When one frame is missing, this absent frame will be filled up with the latest previous frame in our experiment. We compare the performance of multi-modal model, to lip-only and audio-visual model when randomly dropping out 0\%, 10\%, 20\% and 50\% frames.

\begin{table}[th]
  \caption{SI-SDR (dB) performances of target separation models with different dropout rates on lip video frames.}
  \label{tab:v_drop}
  \centering
  \begin{tabular}{l|cc|cc|cc}
    \toprule
    \multirow{2}{*}{\textbf{dropout}} &
    \multicolumn{2}{c|}{\textbf{lip-only}} &
    \multicolumn{2}{c|}{\textbf{audio-visual}} &
    \multicolumn{2}{c}{\textbf{multi-modal}} \\
    & 1spk & ave & 1spk & ave & 1spk & ave  \\
    \hline
 0\%  & 24.0 & 16.2 & 24.7 & 17.1 & 24.8 & 17.2  \\
 10\%  & 23.9 & 15.8 & 24.6 & 17.1 & 24.7 & 17.1 \\
 20\%  & 23.7 & 15.7 & 24.6 & 17.1 & 24.7 & 17.1 \\
 50\% & 23.2 & 15.0 & 24.5 & 17.0 & 24.5 & 17.0 \\
    \bottomrule
  \end{tabular}
\end{table}

Results are presented in Table \ref{tab:v_drop}. For lip-only model, the dropping of frames have an obvious negative effect on the overall performance. While for models integrated with other complementary modalities, the negative influence is alleviated. Especially for the multi-modal model, the performance decrease is less than 2\% when existing 50\% frame drops. This confirms the robustness of our multi-modal model to the missing of visual information.

\subsubsection{Impact of sound direction estimation error}
\begin{figure}[ht]
    \centering
    \includegraphics[width=\linewidth]{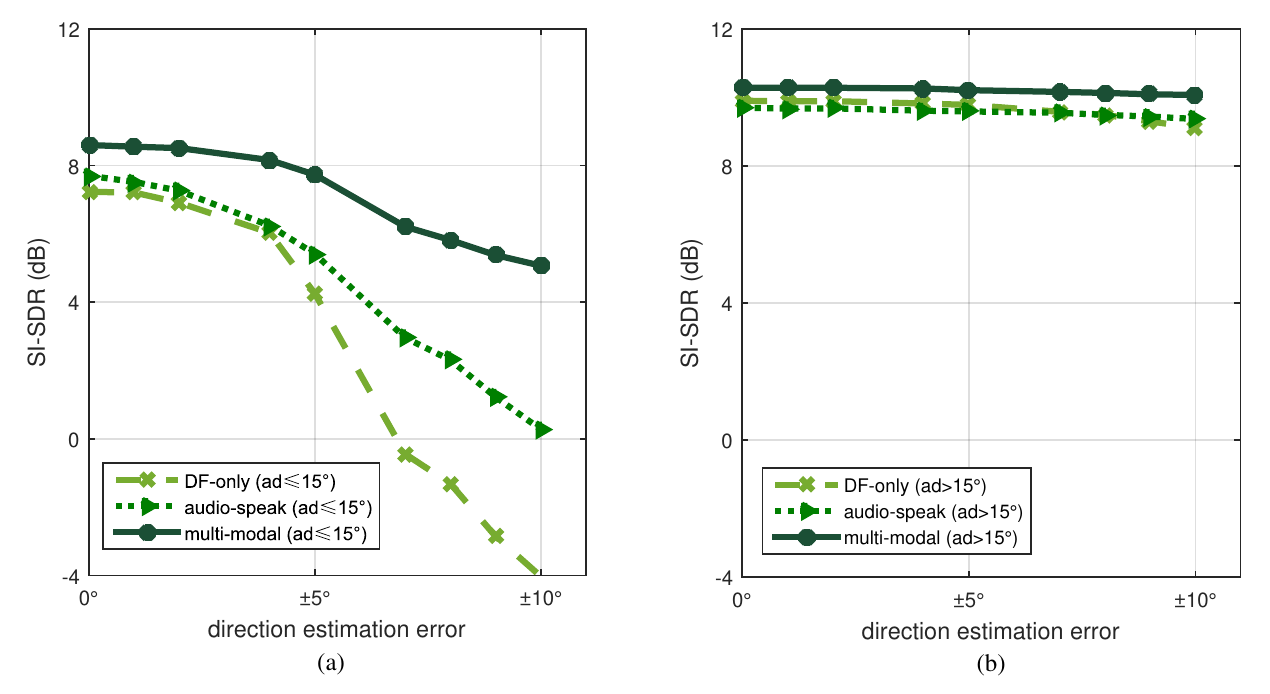}
    \caption{The SI-SDR (dB) performance of audio-only, audio-speaker and multi-modal model when existing the direction estimation error, evaluated on overlapped data. }
    \label{fig:dee}
\end{figure}

For the audio-only model that greatly depends on the directional features, tiny direction estimation error may cause huge estimation inaccuracy. When other modalities are available, the deviation can be remedied to some extent.
We compare the performance of multi-modal model, to DF-only and audio-visual model when there exists an estimation error of the direction detected by the speaker's face. The ground truth direction is deviated for $\pm$1-$\pm$10$\degree$ to compute the target speaker's DF. The performance is examined under two cases: the closest angle difference between target and interfering speaker is smaller than 15$\degree$ and larger than 15$\degree$. Figure \ref{fig:dee} plots the changing curve of performances versus direction estimation errors for three models: DF-only, audio-speaker and multi-modal model.

As we can observe from Figure \ref{fig:dee} (b), fortunately, the performances of all the models under $ad>15\degree$ are robust to the direction estimation error. However, as the direction estimation error increases, the performance of DF-only model degrades dramatically when the target and interfering speaker(s) are close (Figure \ref{fig:dee} (a)). This is due to the spatial ambiguity issue when directional information is not sufficient enough to discriminate between the target and interfering speaker. Since only directional information is served as the target information, the network cannot identify which speaker should be separated. When speaker embedding is integrated into the model (audio-speaker), the dropping of performance relatively slows down. This is because the voice characteristics of the target speaker can complement the target information. Furthermore, when all the target information is aggregated in one single model (multi-modal), the overall performance degradation is less than 1.5dB for the direction estimation error of $\pm 10\degree$.

Experimental results suggest that our proposed multi-modal model exhibits more stable and persistent performance under interferences from video or audio modality.

\section{Conclusion}
In this work, we propose the first deep multi-modal framework for multi-channel target speech separation. The multi-modal framework exploits all sorts of target-related information, including the target's spatial location, lip movements and voice characteristics. Efficient and robust multi-modal fusion approaches are proposed and investigated within the framework. Evaluation on a large-scale audio-visual to-be-released dataset demonstrates the effectiveness and steadiness of the proposed multi-modal system.

This work still has some limitations that needs to be addressed in our future work. Firstly, the joint training of video and audio stream may not produce lip embeddings that are discriminative enough. We will follow the work of \cite{ephrat2018looking} and \cite{afouras2018conversation} to pretrain the lipnet with phonetic transcribed data.
Secondly, although the proposed multi-model system has demonstrated its robustness to error/missing of some of input modalities, data augmentation schemes can be further used to improve the robustness.
Thirdly, the fusion methods investigated in this work are useful but we believe there is still room for improvement.


%

\section*{Acknowledgement}
This research is partly supported by Shenzhen Science \& Technology Fundamental Research Programs (No: JCYJ20170817160058246 \& No: JCYJ20180507182908274).

\ifCLASSOPTIONcaptionsoff
  \newpage
\fi



%

\bibliographystyle{IEEEtran}
\bibliography{mybib}

\end{document}